\documentclass[12pt]{article}
\usepackage{axodraw,epsfig}
\title{\bf The electric charge and magnetic moment of neutral fundamental
 particles}
\bigskip
\author{Kaushik Bhattacharya
\thanks{e-mail
kaushikb@iitk.ac.in 
}\\
\normalsize
Department of Physics,\\
\normalsize 
Indian Institute of Technology, Kanpur,\\
\normalsize
Kanpur 208016, Uttar Pradesh, India.
}
\begin{document}
\maketitle
\begin{abstract}
  The article focuses on the issue of the two definitions of charge, mainly
  the gauge charge and the effective charge of fundamental particles. Most
  textbooks on classical electromagnetism and quantum field theory only works
  with the gauge charges while the concept of the induced charge remains
  unattended. In this article it has been shown that for intrinsically charged
  particles both of the charges remain the same but there can be situations
  where an electrically neutral particle picks up some electrical charge from
  its plasma surrounding. The physical origin and the scope of application of
  the induced charge concept has been briefly discussed in the article.
\end{abstract}
\section{Introduction}
The electric charge of a fundamental particle is inherently related to an
underlying U(1) gauge symmetry of a theory. On the other hand the
electromagnetic properties of an extended charge distribution in an arbitrary
region of space-time has nothing to do with any gauge symmetry. 
Classically the electric and magnetic properties of the extended charge
distribution is specified by the multipole expansion of the scalar and the
vector potential. In quantum field theory the role of the multipole expansion
of the potentials is taken by the form-factors of the electromagnetic vertex
of the charge distribution.

The interesting point is that we can use the process involved to specify the
various moments or form-factors of an unknown extended charge distribution to
specify the electromagnetic properties of the fundamental particles
themselves. It may seem that the electromagnetic form-factors or the multipole
moments of a fundamental particle is an useless concept because in these cases
we know their U(1) gauge charges and more over we are unsure about their
spatial size. But a closer look into the quantum field theoretical foundations
of particle physics tells us that although the facts stated before remain true
but still they are not enough to specify the charge of the fundamental
particles in an interacting theory of quantum fields. The primary reason why
form-factors can play an important role in field theories is related to the
virtual degrees of freedom. The fundamental particle which is real can always
be accompanied by a lot of virtual charged particles and consequently the
charge seen by a photon is effectively modified. The set of virtual charged
particles near the real particle whose charge is in question gives an extended
charge distribution and consequently the multipole moment analysis again
becomes meaningful. As an virtually extended charge system emerges near the
real on-shell particle, the multipole moments become scale dependent or in
other words the electromagnetic form-factors are functions which depend upon
the momentum of the probing photon. Out of all the form factors there is one
which corresponds to the net charge of the system. If the fundamental particle
had a U(1) gauge charge then it definitely was independent of momentum and
consequently in the interacting quantum field theory there must be one
form-factor which reduces to the U(1) charge of the fundamental particle in
the limit of zero momentum. This form-factor is called is called the charge
form-factor. Similarly it may turn up that the fundamental particle has other
form-factors corresponding to magnetic moment, electric dipole moment and
other esoteric moments never encountered in classical electromagnetism. If it
turns out that the fundamental particle in question does not possess any U(1)
gauge charge to start with then it happens that the charge form-factor
vanishes in the zero momentum limit giving a consistent meaning of the charge
of the fundamental particle.

The previous discussion of quantum field theory was primarily done without any
mention of a background plasma of relativistic particles.  In presence of such
a thermal medium new physics emerges. To give one simple example which is
related to the topic of this article is the emergence of screening effect. In
presence of a plasma the electric charge of a charged particle gets screened
by the opposite charges present. The oppositely charged particles will form an
real extended charge system and consequently the effective charge of the
fundamental particle gets modified and this modification is dependent on the
momentum of the probing photon. From the above statements it can be inferred
that the electromagnetic form-factors of a fundamental particle gets modified
from its vacuum values in presence of a background plasma. One interesting
feature of the thermal backgrounds is related to the emergence of an effective
charge of a neutral particle. A neutral fundamental particle may have
electromagnetic (as the photon) or non-electromagnetic (as the neutrino)
interactions with the charged particles in the plasma and consequently it can
produce a charge polarization around it. In this case as we start to probe the
fundamental particle from a distance we will observe a certain amount of
charges of the plasma to be in its sphere of influence and if our probing
region increases indefinitely then there appears a finite number of charged
particles inside the greatest sphere of influence of the fundamental particle.
If the plasma is charge symmetric and the interaction of the particle in
question is also charge symmetric then there will be an equal number of
opposite charged particles polarized around the neutral particle and
consequently the effective charge of the neutral particle vanishes. On the
other hand if the the plasma is not charge-symmetric or the interaction of the
fundamental particle breaks charge conjugation symmetry there is a distinct
possibility that there will be a net charge built up around the neutral
particle.  Consequently we will get a non-zero effective charge.  This is a
kind of inverted-screening effect, where the original charge which was to be
screened is actually nonexistent. In technical language we have then an
effective charge of the fundamental neutral particle in the zero momentum
limit although its U(1) gauge charge is precisely zero.

From our previous discussions we see that we can have two definitions of
charge of a fundamental particle. One is related to the gauge invariance of
the particle under some U(1) gauge group and the other related to its
effective coupling to photons. Both of these definitions agree for a
intrinsically charged particle but in nature we have electrically neutral
particles also and in those cases they may pick up some charge in a plasma.
Books on applications of quantum field theory generally do not shed much light
on the concept of the two definitions of charges, their physical origin and
scope of applications, except some as in Refs.~\cite{Raffelt:1996wa,
  Mohapatra:1998rq}. In research level literature we find few calculations of
the induced charge \cite{Altherr:1993hb, Nieves:1993er} of a neutral particle
in a medium.  Consequently many questions remain unanswered regarding the
application of electrodynamics of neutral fundamental particles. In this
article we will discuss more quantitatively what has been qualitatively
discussed above and try to point out the wider scope and limits of application
of the effective charge concept.

The material in the article is presented in the following manner. In
section \ref{elec} we distinguish between the concepts of the intrinsic or the
gauge charge and the effective charge of the electrons. In section \ref{emneu}
we present a brief outline showing the emergence of an effective electric
charge for the otherwise neutral standard model neutrinos. Section \ref{embos}
contains a brief discussion on the effective induced charge of the photons
and real scalars. The main points in the article are summarized in section 
\ref{dissum}. 
\section{The case of the electrons}
\label{elec}
\subsection{The gauge charge}
Generally the electric charge of any elementary particle is a concept
which is intimately linked with the local gauge invariance of the
theory under an Abelian gauge transformation. In QED if the
electromagnetic gauge field is represented by $A^{\mu}(x)$ then under
a local gauge transformation
\begin{eqnarray}
A^\mu (x) &\to& A^\mu (x)  + \partial_\mu \chi(x)\,,\\
\psi (x) &\to& \psi(x)\,e^{-ieQ\chi(x)}\,,
\label{gt}
\end{eqnarray}
where $\chi(x)$ is a well behaved function of the space-time coordinates and
$e$ is the unit of charge, chosen as charge of the proton, and $Q$ is a
constant designating the amount of charge. The physical meaning of $eQ$ will
become transparent when we write the conserved Noether charge corresponding to
the continuous symmetry of the system. Due to gauge invariance the Dirac
equation for a charged fermion becomes:
\begin{eqnarray}
(i{\rlap /\partial} - eQ {\rlap /A} - m)\psi(x)=0\,.
\label{dirac}
\end{eqnarray}
Using the above equation we can solve for the wave-function for various cases
of the electromagnetic potential $A_\mu$. In particular $A_\mu$ may be a
classical field and chosen in such a way that it produces a constant electric
or magnetic field or both. In some particular cases we can exactly solve
Eq.~(\ref{dirac}) for electric or magnetic field backgrounds.  The solutions
portray the behavior of a charged fermion with electric charge $eQ$
interacting with the classical electric or a classical magnetic field.  If we
solve Eq.~(\ref{dirac}) in presence of such an $A_\mu$ which produces a
uniform magnetic field we will get exact non-perturbative (in the external
magnetic field strength) solutions whose energy in the transverse direction of
the field is Landau quantized. In this case we see that the quantum charged
particle interacts with the magnetic field and does something which is
analogous to the motion of a classical charged particle whose trajectory is
predicted by the Lorentz force law.

In the non-relativistic limit Eq.~(\ref{dirac}) reduces to the
Schrodinger-Pauli 2-component equation
\begin{eqnarray}
\left[\frac{1}{2m}\left({\bf p} - eQ{\bf A}\right)^2 - \frac{eQ}{2m}
\mbox{\boldmath$\sigma$} \cdot {\bf B} + eQ \phi
\right]\psi_A = E\psi_A \,, 
\label{spauli}
\end{eqnarray}
where the Dirac spinor is written as
\begin{eqnarray}
\psi = e^{-iEt}
\left(
\begin{array}{c}
\psi_A\\
\psi_B
\end{array}
\right)\,,
\end{eqnarray}
and the classical vector potential giving rise to the magnetic field ${\rm B}$
is $A^\mu=(\phi, {\bf A})$. From the above equation it is clear that the
magnetic moment of the electron is $\frac{eQ}{2m}\mbox{\boldmath$\sigma$}$.
The point to note is that, the charge $eQ$ is the unique coupling constant
which couples the Dirac particle to the electric and the magnetic field and
this coupling is prescribed by gauge invariance of the theory.

In quantum field theory the conserved Noether current corresponding to
an infinitesimal gauge transformation as given in Eq.~(\ref{gt}) is
\begin{eqnarray}
J_{\rm N}^\mu = eQ\,\overline{\psi}\,\gamma^\mu\,\psi\,.
\label{n_curr}
\end{eqnarray}
Using the Fourier expansion of the free fermionic fields we find that
the conserved charge\footnote{Generally Noether´s charge is defined
for a global symmetry but in this case we can also define it for the
local gauge invariance as the corresponding form of the Noether charge
in QED is similar in form to the charge as defined in the case of
global symmetries. The topic is discussed in
Ref.~\cite{AlKuwari:1990db}.} for the Abelian gauge symmetry is
\begin{eqnarray}
Q_N &=&\int d^3 x \,J_{\rm N}^0 = eQ \int d^3 x \,
\psi^\dagger(x)\psi(x)\,,\\
&=& eQ \int d^3 p\,\sum_{s=1,2}\left[a^\dagger_s({\bf p}) a_s({\bf p})
- b^\dagger_s({\bf p})b_s({\bf p}) \right]\,.
\end{eqnarray}
Here $ a_s({\bf p})$ and $b_s({\bf p})$ are the annihilation operators
for a free electron field and the free positron field
respectively. The last equation implies that $eQ$ is indeed related to
the charge of the fields. For one single electron $Q=-1$ and for a
positron $Q=1$.  Conventionally $eQ$ is called the charge of the field
$\psi$. If the field $\psi$ does not transform under a gauge
transformation, is a gauge singlet, then $Q=0$ and we do not have any
interaction of the fermion with the electromagnetic field. In quantum
field theory all the particles are excitations of the vacuum, which
are created or destroyed by the relevant creation and destruction
operators. From the above analysis we see that all these particles
have a charge which is given by the eigenvalue of the charge operator
$Q_N$ acting on the relevant particle states. The charge of the
electron which entered the Dirac equation via minimal prescription is
the eigenvalue of $Q_N$ when the charge operator acts on a free
electron momentum state.
\subsection{The effective charge}
In quantum field theory, the electric charge of fermions can also be
understood in an {\it effective} way. In this effective scheme it is
assumed that the interaction of any off-shell photon with two on-shell
fermions is of the form
\begin{eqnarray}
-J^\mu  (x)\,A_\mu (x)\,,
\label{currfield}
\end{eqnarray}
where $J^\mu (x)$ contains all the information about the interaction
of the on-shell fermions. The matrix element of the above current is
conventionally written as
\begin{eqnarray}
\langle e^-(p^\prime,{s^\prime})|J^\mu(x)|e^-(p,s)\rangle = 
\frac{e^{-iq\cdot x}}{\sqrt{2E_p V}\sqrt{2E_{p^\prime} V}}\overline
{u}_{s^\prime}({\bf p}^\prime)\,e\Gamma_\mu (p,p^\prime)\,u_s({\bf p})\,,
\label{effcurr}
\end{eqnarray}
where now $\Gamma_\mu (p,p^\prime)$ is an effective 4-vector vertex which can
depend only upon the 4-moment's of the electrons in the current, $V$ is some
volume element and $u_s({\bf p})$ is the Dirac spinor with momentum ${\bf p}$
and spin $s$. Using the condition of electromagnetic current conservation we
can write the most general form of $\Gamma_\mu (p,p^\prime)$ as:
\begin{eqnarray}
\Gamma_\mu (p,p^\prime) = \gamma_\mu F_1(q^2)+\left(iF_2(q^2) + 
F_3(q^2)\gamma_5\right)\sigma_{\mu\nu}q^\nu + F_4(q^2)\left(q_\mu {\rlap q/} - 
q^2\gamma_\mu\right)\gamma_5\,,
\label{qedvert}
\end{eqnarray}
where $q=p-p^\prime$ and $F_1(q^2)$ is called the {\it charge form-factor} of
the electron, $F_2(q^2)$ is related to the {\it anomalous magnetic moment} of
the electron, $F_3(q^2)$ is related to the {\it electric dipole moment} of the
electron and $F_4(q^2)$ is called the {\it anapole moment} of the electron. In
the last equation
\begin{eqnarray}
\sigma_{\mu \nu}=\frac{i}{2}\left[\gamma_\mu\,,\,\gamma_\nu\right]\,.
\end{eqnarray}
In this article we will only concentrate on the charge form-factor and
the magnetic moment form-factors leaving out the other form-factors
which are excellently discussed in \cite{Lahiri:2005sm}. For the case
when the initial and the final electrons have the same spin and same
4-momentum i.e., $s=s^\prime$ and $p=p^\prime$ we can define an
effective charge of the electron from Eq.~(\ref{effcurr}) and
Eq.~(\ref{qedvert}) as
\begin{eqnarray}
Q_{E}=\frac{1}{2E_p}\overline{u}_{s}({\bf p})
\Gamma_0(q_0=0,\,{\bf q} \to 0) 
  \,u_s({\bf p})\,.
\label{chargedef}
\end{eqnarray}
Substituting the above result in Eq.~(\ref{effcurr}) and then comparing with
Eq.~(\ref{n_curr}) yields $Q_{E}=F_1(0)=Q$. The last equation serves as an
effective definition of the charge of an electron and is independent of the
definition of the charge as predicted from the Noether current, although in
the present case both the charges are the same. The limiting procedure of
taking $q_0=0$ and then ${\bf q}\to 0$ is some times called the {\it static
  limit}. In the case above the static limit is irrelevant, in the sense that
we could also have worked with the limiting prescription ${\bf q}=0$ and then
$q_0 \to 0$, as long as Lorentz symmetry is not violated.

Now if we choose an $A^\mu (x) = (0, {\bf A}(x))$ and take
non-relativistic fermions, it can be shown that the interaction term
\begin{eqnarray}
-j_\mu A^\mu  \propto \frac{eF_1(0)}{2m}\mbox{\boldmath$\sigma$} 
\cdot {\bf B}
\label{nmm}
\end{eqnarray}
where we have taken the limit $q^2 \to 0$ and ${\bf B}$ is the classical
magnetic field corresponding to ${\bf A}(x)$. From the above
expression we see that if $F_1(0)=Q$ we get back the normal magnetic moment
\begin{eqnarray}
\mbox{\boldmath $\mu$}= \frac{eQ}{2m}\mbox{\boldmath $\sigma$}\,,
\end{eqnarray}
of the electron as predicted from Eq.~(\ref{spauli}). So we see that the the
effective charge and the gauge charge of the electrons behaves similarly 
as far as their interaction with electric or magnetic fields are
concerned. 

The anomalous magnetic moment term also gives a similar contribution as above
except that it does not depend on $Q=F_1(0)$ but on $F_2(0)$. Consequently an
uncharged particle can also have an anomalous magnetic moment.  While it is
obvious from the anomalous magnetic moment interaction term that it
contains $\sigma_{\mu\nu}$ it can also be shown by Gordon identity that the
normal magnetic moment interaction, as appearing in Eq.~(\ref{nmm}), was
obtained from a term containing $\sigma_{\mu\nu}$. As
\begin{eqnarray}
\overline{\psi}\sigma_{\mu \nu} \psi = \overline{\psi}_R \sigma_{\mu \nu} 
\psi_L + \overline{\psi}_L\sigma_{\mu \nu} \psi_R
\end{eqnarray}
where $\psi_L= \frac12(1-\gamma_5)\psi$ and $\psi_R= \frac12(1+\gamma_5)\psi$
are the left-handed and right-handed fermion field projections. Consequently
we can say that if due to some reason only $\psi_L$ or $\psi_R$ is present
then there will be no magnetic moment of the fermions as the magnetic moment
is a chirality flipping operator.
\section{Electromagnetic interaction of the neutrinos}
\label{emneu}
In the symmetry broken phase of the standard model the neutrinos do not have
any electric charge, consequently $Q=0$ for them. More over there are only
left-handed neutrinos and right-handed antineutrinos in the present universe.
The Dirac equation for the neutrino is:
\begin{eqnarray}
i{\rlap /\partial}\,\psi_L(x)=0\,.
\label{diracn}
\end{eqnarray}
As the gauge charge of this neutrino is zero the neutrino does not feel the
electric or magnetic field, as an electron feels, and consequently its motion
in the transverse plane of a magnetic field does not get quantized. But if we
conclude from the last equation that the neutrino can never feel any electric
or magnetic field then the conclusion will be wrong. The reason for the error
is that the neutrino may not have a gauge coupling with the electromagnetic
fields but still we have not discussed about its effective coupling with the
electromagnetic field. From the interactions in the Standard Model of particle
physics we know that the neutrinos can interact with the electrons and other
charged particles (which have a proper gauge charge associated with them)
through the charged-currents, and those charged particles can interact with
the photons directly. So the neutrinos can interact with $A_\mu$ in an
indirect way. This indirect interaction can also be described by
Eq.~(\ref{currfield}) where now the current $J^\mu$ corresponds to the
neutrino current.
\subsection{The neutrino-photon effective coupling}
Not repeating all the points stated before for electrons we can directly use
the formula in Eq.~(\ref{chargedef}) for the effective charge of the neutrino
with a minor modification. The modification of Eq.~(\ref{chargedef}) is
related to the handedness of the Standard Model neutrinos. Now the two spinors
sandwiching $\Gamma_0(q_0=0,\,{\bf q} \to 0)$ in Eq.~(\ref{chargedef}) will
be replaced by the left-handed spinors, corresponding to the left-handed
neutrinos. The modified equation for the neutrino effective charge becomes
\begin{eqnarray}
Q_\nu = \frac{1}{2E_p}\overline{u}_{L}({\bf p})
\Gamma_0(q_0=0,\,{\bf q} \to 0) 
  \,u_L({\bf p})\,. 
\label{nchargedef}
\end{eqnarray}
%
\begin{figure}[t]
\begin{center}
\begin{picture}(100,100)(-50,-20)
\ArrowLine(80,-10)(0,0) 
\Text(35,-15)[c]{$\nu(p)$} 
\ArrowLine(0,0)(-80,-10)
\Text(-35,-15)[c]{$\nu(p')$}
\Text(0,52)[c]{$\mu$} 
\ArrowArc(0,30)(30,90,270)
\Text(-38,30)[c]{$\ell$}
\ArrowArc(0,30)(30,-90,90)
\Text(0,7)[c]{$\alpha$}
\Photon(0,60)(0,100)26
\ArrowLine(-7,70)(-7,85)
\Text(5,80)[l]{$\gamma(q)$}
\end{picture}
\caption[]{\sf The effective neutrino electromagnetic vertex in the
leading order in the Fermi constant $G_F$ and the electric charge of
the lepton $q_{\ell}$ inside the loop. The indices $\alpha$ and $\mu$
designate the 4-Fermi and QED couplings of the neutrinos with the
leptons and the leptons with the photons respectively.
\label{f:4fermi}}
\end{center}
\end{figure}
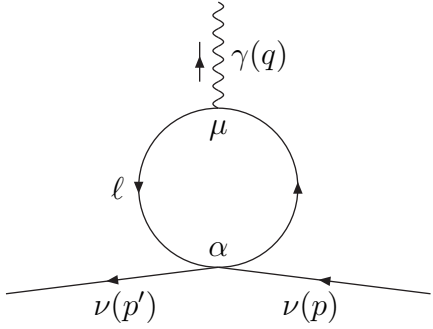
But to apply the equation of the effective charge of the neutrinos we first
require a form of the effective neutrino-photon vertex function $\Gamma_\mu$.
From Hermiticity, electromagnetic current conservation principle and
\begin{eqnarray}
q^\mu \Gamma_\mu = 0\,,
\label{qgamma0}
\end{eqnarray}
which is the Ward identity for a particle which lacks any kind of
gauge charge attached to it we can write the most general form of the
effective vertex for the chiral neutrinos as \cite{Nieves:1981zt}:
\begin{eqnarray}
\Gamma_\mu (p, p^\prime)= \left(q^2 \gamma_\mu - q_\mu {\rlap /q} \right)
\left(R(q^2) + r(q^2)\gamma_5\right)\,,
\label{effvertn}
\end{eqnarray}
where $R(q^2)$ and $r(q^2)$ are real form-factors often called the {\it charge
  radius} and the {\it axial charge radius} respectively. The important point
now is to see whether $R(q^2)$ and $r(q^2)$ in $\Gamma_\mu$ has any poles as
$q^\mu \to 0$, as only then there is a possibility of a finite $Q_\nu$. The
form factors are obtained by calculations of loops which are inscribed in
between two on-shell neutrinos and an off-shell photon as shown in
Fig~\ref{f:4fermi}. In any Feynman diagram which contains an off-shell photon
which is effectively coupled to two on-shell neutrinos the photon must couple
to two internal lines of charged particles. If one of the lines is assigned a
loop momentum $k$ then the other must carry momentum $k\pm q$. The propagator
of the second line will involve the factor
\begin{eqnarray}
\frac{1}{(k \pm q)^2 - m_{\ell}^2}=
\frac{1}{q^2 \pm 2k\cdot q + (k^2 - m_{\ell}^2)}\,,
\end{eqnarray}
where $m_{\ell}$ is the mass of the particle in the internal
line. As for any internal line $k^2 \ne m_{\ell}^2$, no singularity
is present in the expressions of the form-factors. Consequently
$\Gamma_0(q_0=0, {\bf q}\to 0) = 0$ and from Eq.~(\ref{nchargedef}) we see
that there is no effective neutrino charge in this case. More over as
because there are no right-handed neutrinos in the Standard Model of
particle physics there isn't any magnetic moment of the neutrinos. So a
Standard Model neutrino does not have a gauge charge, does not have an
effective charge, does not have any normal or anomalous magnetic
moment. Classically any particle having such properties will never
interact with any electromagnetic field but in quantum field theory it
is not so. As $\Gamma_\mu (p, p^\prime)$ is in general not zero for
non-zero $q$ so the effective interaction of the neutrino with the
electromagnetic fields is not zero. 
\subsection{The neutrino-photon effective coupling in a thermal background}
As discussed in the introduction we know that the vertex function in presence
of a background plasma has new physics in it and so heceforth in this article
we will designate the neutrino-photon effective vertex in the plasma as
$\Gamma_\mu^\prime (p, p^\prime)$.  The effective neutrino-photon vertex
$\Gamma_\mu^\prime (p, p^\prime)$ in this case also follows all the
constraints as proposed in the last subsection as Hermiticity, current
conservation and the Ward identity.  The definition of effective charge of the
chiral neutrino in the presence of a medium can also be written as in
Eq.~(\ref{nchargedef}), but the static limit of the vertex function has a
subtle preference here.  The reason for the preference for the static limit
here is related to the fact that the temperature and the chemical potential of
the plasma constituents are well defined only in the rest frame of the medium
and so the calculation of $\Gamma_0^\prime$ must be done in an unique frame.
This breaks the Lorentz invariance of the theory. In absence of Lorentz
symmetry the limiting process $q \to 0$ looses meaning because it can be taken
in two ways as first $q_0=0$ and then taking ${\bf q}\to 0$ or taking ${\bf
  q}= 0$ first and then $q_0 \to 0$ and these two ways produces different
results in a thermal medium.

Assuming that $\Gamma_\mu^\prime$ is analytic or at best it has a
removable singularity at $q^0=p^0-p^{\prime 0}=0$ we can Taylor expand 
$\Gamma_\mu^\prime$ about the point $q^0=0$ as \cite{Nieves:1993er}:
\begin{eqnarray}
\Gamma_0^\prime = G_0 + q^0 G_1 + O\left((q^0)^2\right)\,,\\
{\bf \Gamma}^\prime = {\bf H}_0 + q^0 {\bf H}_1 + O\left((q^0)^2\right)\,,
\end{eqnarray} 
where the coefficients of the expansion are well behaved at $q^0=0$.
As Eq.~(\ref{qgamma0}) still holds we have,
\begin{eqnarray}
q_0 G_0 + q_0^2 G_1 =  {\bf q}\cdot{\bf H}_0 + q_0 {\bf q}\cdot{\bf H}_1\,.
\nonumber
\end{eqnarray}
When $q^0 \to 0$ the above relation becomes $q_0(G_0 - {\bf
q}\cdot{\bf H}_1) - {\bf q}\cdot{\bf H}_0 = 0$ and as $q_0$ and ${\bf
q}$ are independent quantities for an off-shell photon, the last
equation implies $G_0 = {\bf q}\cdot{\bf H}_1$ and ${\bf H}_0 =
0$. Using these equations in the expansion of $\Gamma_0^\prime$
and ${\bf \Gamma}^\prime$ as $q^0 \to 0$ we get
\begin{eqnarray}
\Gamma_0^\prime &=& {\bf q}\cdot{\bf H}_1 + O(q^0)\,,\\
{\bf \Gamma}^\prime &=& q^0 {\bf H}_1 + O\left((q^0)^2\right)\,.
\end{eqnarray}
Inside a medium the term ${\bf H}_1$ can always have a pole at ${\bf q}=0$
as a medium consists of real particles and they may enter the loop integrals
contributing for the calculation of $\Gamma_0^\prime=0$. For these looping
real particles with 4-momentum $k^\mu$, we may have $k^2 = m_{\ell}^2$ unlike
the case in vacuum. Consequently if ${\bf H}_1$ develops a pole at ${\bf
  q}=0$, where its form is given as
\begin{eqnarray}
{\bf H}_1 \propto \frac{{\bf q}}{|{\bf q}|^2}\,,
\end{eqnarray}
then $\Gamma_0^\prime$ will not be zero and will become a finite quantity as
$q^0 \to 0, {\bf q}\to 0$. In a medium this happens and consequently the
neutrino develops an effective electric charge.  If the effective 4-Fermi
interaction is given as,
\begin{eqnarray}
{\mathcal L}_{\rm eff} = 
-\surd2 G_F \Big[\overline\psi_{(\nu)}\gamma^\alpha L\psi_{(\nu)} \Big]
\Big[\overline\psi_{(\ell)} \gamma_\alpha (g_V + g_A\gamma_5)
\psi_{(\ell)} \Big] \,,
\label{4fermi}
\end{eqnarray}
where $L = \frac12(1 - \gamma_5)$ is the left handed projection operator, then
from Fig.~\ref{f:4fermi}, the effective neutrino-photon vertex to
one-loop level can be written as
\begin{eqnarray}
\Gamma_\mu = - \, {\surd2 G_F \over q_\ell} \gamma^\alpha L \Big( g_V
\Pi_{\mu\alpha} + g_A \Pi^5_{\mu\alpha} \Big) \,.
\label{nvf}
\end{eqnarray}
Here $g_V$ and $g_A$ are the vector and axial-vector couplings of the
neutrinos to the leptons and $q_\ell$ is the electric charge of the looping
lepton.  The term $\Pi_{\mu\alpha}$ is exactly the expression for the vacuum
polarization of the photon, and appears from the vector interaction in the
effective Lagrangian. In the one-loop expression of $\Pi_{\mu\alpha}$ if we
replace one of the vector vertices by an axial-vector vertex then we get
$\Pi^5_{\mu\alpha}$. $\Pi_{\mu\alpha}$ and $\Pi^5_{\mu\alpha}$ arises from the
vector and the axial-vector coupling of the neutrinos to charged leptons in
the 4-Fermi Lagrangian. In Fig.~\ref{f:4fermi} the vertex specified by the
Greek letter $\alpha$ is the 4-Fermi vertex and the Greek letter $\mu$ stands
for the vector vertex. The Feynman diagram explains the origin of the two
polarization tensors.  To one-loop order an exact calculation of the neutrino
effective charge in a medium gives \cite{Nieves:1993er}:
\begin{eqnarray}
Q_\nu = -\frac{\sqrt{2} G_F g_V}{q_\ell} \Pi_{00}\,.
\end{eqnarray}
This result shows us that in a medium there is indeed an effective electric
charge of the neutrinos. In the absence of any right-handed neutrinos, the
magnetic moment of the neutrino still remains zero in presence of a medium.
It is to be noted that the form of the vertex function as given in
Eq.~(\ref{nvf}) is a one-loop result which also holds in vacuum. The vertex
function written in Eq.~(\ref{effvertn}) holds only in vacuum but it is the
most general form of the electromagnetic vertex function of neutrinos and
holds for any order of perturbation theory.
\section{A brief discussion on the electromagnetic interactions of neutral 
bosons}
\label{embos}
In quantum field theory photons are the Abelian local gauge fields and do not
carry any gauge charge. In other words photons do not interact with photons
directly. But this statement does not rule out the effective coupling of
photons with photons. As for the case of electrons and neutrinos here also we
can write the effective electromagnetic current of the photons which interacts
with another off-shell photon via intermediate virtual charged
particles.  The matrix element of the electromagnetic current of the photon,
in vacuum, can be written analogously as
\begin{eqnarray}
\langle A^{\prime \alpha^\prime}(p^\prime, \epsilon^\prime)|
[J_\mu(x)]_{\alpha^\prime \alpha}|A^{\alpha}(p,
\epsilon)\rangle= \frac{e^{-iq\cdot x}}{\sqrt{2E_p V}\sqrt{2E_{p^\prime} V}}
\epsilon^{\prime\,*\,\alpha^\prime}(p^\prime)\epsilon^\alpha (p)
\Gamma_{\alpha^\prime \alpha \mu}(p,p^\prime)
\end{eqnarray}
where $\epsilon^{\prime\,\alpha^\prime}(p^\prime)$ and $\epsilon^\alpha(p)$
are the polarization vectors of the on-shell photons with energies
$E_{p^\prime}$ and $E_p$. In the above equation $V$ is the volume element in
which the events occur and in the end of the calculation we can take it to be
infinite. The momentum change of the on-shell photons is given by $q=p^\prime
- p$. In the above equation $\Gamma_{\alpha^\prime \alpha \mu}(p,p^\prime)$ is
the effective vertex function of three photons. The expression of the
electromagnetic vertex of neutral spin-1 particles shades some light on the
electromagnetic interaction of photons in vacuum , and this topic has been
discussed in general in Ref.~\cite{Nieves:1996ff}. Without doing any explicit
calculation it can be said that in QED this effective vertex vanishes as any
charged fermion loop with odd number of photon vertices vanishes due to charge
conjugation symmetry, a statement formally known as Furry's theorem in QED.
Consequently in vacuum odd number of photons do not interact with themselves
in any order of perturbations, unless we include fields which breaks charge
conjugation symmetry as neutrinos. But the situation changes if we have a
background thermal medium which breaks charge symmetry in a trivial way. One
of the simplest example is the plasma atmosphere of stars or planets whose
temperature is less than $0.5{\rm MeV}$. In these kind of atmospheres there
will be more electrons than positrons. In this case the photon electromagnetic
effective vertex does not vanish and odd number of photons start to interact
with each other in an effective way.  The effective charge of the photon is
related with $\Gamma_{\alpha \, \alpha \,0}(q_0=0,\,{\bf q} \to 0)$ and in
principle it exists. A calculation of the effective charge of the photon has
been done using techniques of plasma physics and statistical mechanics in
Ref.~\cite{Mendonca:2000tk}.

As for the neutral photons the real scalar fields do not have any direct
electromagnetic interactions. But these fields can interact with other charged
scalar fields or can interact with charged fermions via Yukawa
interactions. The charged scalar or fermions will interact with photons giving
an indirect linkage of the real scalars with the electromagnetic field. The
effective vertex of the real scalars with the photons is given as
\begin{eqnarray}
\langle \phi(p^\prime)|J_\mu(x)|\phi(p)\rangle
=\frac{e^{-iq\cdot x}}{\sqrt{2E_p V}\sqrt{2E_{p^\prime} V}} 
\Gamma_\mu(p^\prime,p)\,,
\end{eqnarray}
where all the terms in the above equation has the conventional meaning. Using
the conservation condition of the electromagnetic current, the on-shell
property of the scalar fields the effective electromagnetic vertex of the
scalar field can be written as \cite{Nieves:2007jz}:
\begin{eqnarray}
\Gamma_\mu(p, q) = a[p\cdot q \,q_\mu - q^2 p_\mu]\,,
\end{eqnarray}
where $\Gamma_\mu(p, q)$ has been expressed in terms of $p$ and $q=p^\prime -
p$ instead of $p^\prime$ and $p$ alone and $a$ is a constant. As in the case
of the neutrinos in vacuum in this case also we do not expect any singularities
of $a$ as $q \to 0$. So the above equation implies that $\Gamma_\mu(q_0=0,
{\bf q}\to 0)=0$ and consequently there will be no effective charge of the
real scalar fields in vacuum. In a background medium we can also write the
electromagnetic vertex of the scalars as done in Ref.~\cite{Nieves:2007jz}. In
presence of a background medium the real scalars can in principle acquire an
induced charge analogous to the case of the neutrinos.  
\section{Discussion and summary}
\label{dissum}  
In this article we have shown that the concept of electric charge is not
solely related with Noether's current or an Abelian gauge symmetry. Although
whenever some one speaks about charge an Abelian gauge symmetry naturally
comes to our mind. Precisely for this reason the point about the induced
charge requires some understanding because of the various notions we
unconsciously attach with the {\it gauge charge} of the particles compels us
to think about the induced charge in a wrong way. When ever we find that a
particle is charged we want to understand how does two charged particles
interact with each other? How does the charged particle respond to a classical
electric or a magnetic field? To properly answer these questions we have to
understand the origin of the induced charge. As we see that none of the
neutral elementary particles acquire an effective charge in vacuum
consequently they only aquire it from a medium which is comprised of charged
particles (and antiparticles). Thus the induced charge is a byproduct of a
collective phenomenon. In presence of an external static electric field (where
${\bf E}=-\nabla \phi({\bf x}, t)$) these induced charge acts like a normal
gauge charge and will get repelled or attracted according to the situation.
This happens because the definition of the effective charge presented in this
article is based on the interaction of the net acquired charge and the
electrostatic potential of an external photon.  But the induced charge does
not respond to any classical magnetic field via the Lorentz force law. This is
because in the quantum level the Lorentz force law is an outcome of the gauge
invariance of the theory (where the gauge field comes in the scene through
minimal prescription) and in the case of particles with zero gauge charge
there will be no effect. Similarly the induced charge of the particles does
not impart any dipolar interactions. For dipolar interactions the relevant
quantity required is the electric dipole moment form-factor or the magnetic
dipole moment form-factor of the neutral particle.  To interact with the
dipole moments with an external magnetic or electric field the neutral
particle must acquire a net magnetic or electric dipole moment from the
thermal bath and as long as it does not have that the test particle will not
interact with any field through its dipoles. In the presence of a strong
classical magnetic field the property of the charged particles of the medium
themselves changes and consequently now the charge build up around the neutral
particle will happen in a different way as compared to the charge build up in
the absence of the magnetic field. In presence of an external magnetic field
the charged particles of the medium will get Landau quantized and henceforth
their way of accumulation around, or dispersion from, the neutral particle
will change and consequently the effective charge built up will also change
\cite{Bhattacharya:2001nm}.

In conclusion we can say that the intrinsic charge of a fundamental particle
has multiple roles to play. It appears in the various other
electromagnetic couplings of the particle as in the magnetic moment, electric
dipole moment etc, etc. But the effective induced charge of any fundamental
neutral particle, which only becomes important in presence of a thermal
background of intrinsically charged particles, has a much lower status. To
find out the proper electromagnetic coupling of a neutral particle we have to
calculate the relevant electromagnetic form factors appearing in the effective
electromagnetic vertex and only then can we be certain of the various 
electromagnetic couplings of the neutral elementary particle.

\end{document}